\documentclass[twocolumn,superscriptaddress,prl]{revtex4-1} 

\usepackage{graphicx} 
\usepackage{natbib} 
\usepackage[usenames,dvipsnames]{color} 
\usepackage{amsmath} 
\usepackage{amssymb} 
\usepackage{soul} 
\usepackage{ifthen}
\usepackage{upgreek}
\usepackage[ampersand]{easylist}
\usepackage{todonotes}
\usepackage{verbatim}

\newcommand{\nn}{\nonumber}

\newcommand{\bra}[1]{\langle{#1}|}
\newcommand{\ket}[1]{|{#1}\rangle}

\def\l{\left}
\def\r{\right}
\def\be#1\ee{\begin{equation}#1\end{equation}}
\def\ba#1\ea{\begin{align}#1\end{align}}
\def\bg#1\eg{\begin{gather}#1\end{gather}}
\def\t{\text}

\begin{document}

\title{Implementation of a Walsh-Hadamard gate in a superconducting qutrit}

\author{M.A. Yurtalan}
\email{mayurtalan@uwaterloo.ca}
\affiliation{Institute for Quantum Computing, University
	of Waterloo, Waterloo, ON, Canada N2L 3G1}
\affiliation{Department of Physics and Astronomy,  University
	of Waterloo, Waterloo, ON, Canada N2L 3G1}
\affiliation{Department of Electrical and Computer Engineering,  University
	of Waterloo, Waterloo, ON, Canada N2L 3G1}

\author{J. Shi}
\affiliation{Institute for Quantum Computing, University
	of Waterloo, Waterloo, ON, Canada N2L 3G1}
\affiliation{Department of Physics and Astronomy,  University
	of Waterloo, Waterloo, ON, Canada N2L 3G1}

\author{M. Kononenko}
\affiliation{Institute for Quantum Computing, University
	of Waterloo, Waterloo, ON, Canada N2L 3G1}
\affiliation{Department of Physics and Astronomy,  University
	of Waterloo, Waterloo, ON, Canada N2L 3G1}

\author{A. Lupascu}
\email{alupascu@uwaterloo.ca}
\affiliation{Institute for Quantum Computing, University
	of Waterloo, Waterloo, ON, Canada N2L 3G1}
\affiliation{Department of Physics and Astronomy,  University
	of Waterloo, Waterloo, ON, Canada N2L 3G1}
\affiliation{Waterloo Institute for Nanotechnology, University
of Waterloo, Waterloo, ON, Canada N2L 3G1}

\author{S. Ashhab}
\affiliation{Qatar Environment and Energy Research Institute, Hamad Bin Khalifa University, Qatar Foundation, Qatar}

\date{ \today}

\begin{abstract}
We have implemented a Walsh-Hadamard gate, which performs a quantum Fourier transform, in a superconducting qutrit. The qutrit is encoded in the lowest three energy levels of a capacitively-shunted flux device, operated at the optimal  flux-symmetry point. We use an efficient decomposition of the Walsh-Hadamard gate into two unitaries, generated by off-diagonal and diagonal Hamiltonians respectively. The gate implementation utilizes simultaneous driving of all three transitions between the three pairs of energy levels of the qutrit, one of which is implemented with a two-photon process. The gate has a duration of 35 ns and an average fidelity over a representative set of states, including preparation and tomography errors, of $99.2$\%, characterized with quantum state tomography. Compensation of ac-Stark and Bloch-Siegert shifts  is essential for reaching high gate fidelities.
\end{abstract}
\maketitle
In recent years, significant progress has been made towards the implementation of quantum computers. Current efforts are mainly focused on encoding quantum information using two-state systems, or qubits. Using multi-level systems, or qudits, instead of qubits to perform quantum information processing is a developing field that promises advantages in a number of areas of quantum information. Universal quantum control of qudits and quantum error correction approaches have been explored theoretically~\cite{gottesman_fault-tolerant_1999,grassl_efficient_2003,bullock_asymptotically_2005}. Recent theoretical work suggests that quantum error correction with qudits has potential advantages over qubit-based schemes~\cite{campbell_magic-state_2012,campbell_enhanced_2014,anwar_fast_2014,krishna2019towards}. The experimental implementation of quantum computing based on qudits is still largely unexplored. Besides quantum computing, qudits have been explored as alternatives to qubits in other areas of quantum information, as an improved platform for quantum metrology~\cite{suslov_quantum_2011} and quantum communication~\cite{bouchard_high-dimensional_2017}.

In this Letter, we report the implementation of the generalized Walsh-Hadamard gate in a superconducting three-state qudit, or qutrit. The Walsh-Hadamard gate is one of the elementary gates in qudit control, relevant for error correction~\cite{gottesman_fault-tolerant_1999,grassl_efficient_2003} and the implementation of the quantum Fourier transform in single- and  many-qudit systems~\cite{muthukrishnan_quantum_2002}. We use a fast single-pulse implementation of the gate based on a single rotation in the qutrit space. We note that superconducting devices provide a natural platform for the exploration of the physics of qutrits, with work to date including basic control and tomography~\cite{bianchetti_control_2010}, emulation of spin dynamics~\cite{Neeley722} and topological states of matter~\cite{tanTopologicalMaxwellMetal2018}, the use of the third level of a qutrit to facilitate two-qubit gates~\cite{strauch_quantum_2003}, wave mixing~\cite{honigl-decrinis_mixing_2018}, holonomic gates~\cite{abdumalikov_jr_experimental_2013}, electromagnetic induced transparency~\cite{abdumalikov_electromagnetically_2010}, demonstration of quantum contextuality~\cite{jerger_contextuality_2016}, and adiabatic state transfer protocols~\cite{vepsalainen2020simulating}.

There have been several studies in the literature on decomposing qudit gates into sequences of simple steps, e.g.~into a sequence of gates that each operates on two quantum states~\cite{vitanovSynthesisArbitrarySU2012, klimovQutritQuantumComputer2003,Ramakrishna2000Explicit}.
In this work we implement the Walsh-Hadamard gate for a qutrit, given by   
\begin{equation}
U_\t{WH}=\frac{1}{\sqrt{3}}
\begin{pmatrix}
1&1&1\\
1&e^{i\frac{2\pi}{3}}&e^{-i\frac{2\pi}{3}}\\
1&e^{-i\frac{2\pi}{3}}&e^{i\frac{2\pi}{3}}
\end{pmatrix},
\label{eq:cgate}
\end{equation}
using a decomposition that only requires two steps. Specifically, $U_{\t{WH}}=U_\t{d} U_\t{o}$, with $U_\t{o} =\exp (-iG_\t{o} t) $ and $U_\t{d} =\exp (-iG_\t{d} t) $. The generators $G_\t{o} = \l ( \sum_{0\leq j < k \leq 2} m_{jk} \ket{j} \bra{k} \r) + \emph{h.c.} $ and $G_\t{d} = \t{diag}\l( \phi_0, \phi_1, \phi_2 \r)$. We note that there are multiple distinct decompositions of this type with different values of the complex numbers $m_{01}$, $m_{02}$, and $m_{12} $ and real numbers $\phi_0$, $\phi_1$, and $\phi_2$ and we chose the decomposition that results in the shortest pulse duration for a given drive amplitude. This type of decomposition is well suited  for superconducting circuits where microwave-based control allows for the application of broadband signals containing multiple frequency tones that can simultaneously drive transitions between different levels, allowing to readily implement $G_\t{o}$. The effect of the diagonal unitary $U_\t{d}$ can be implemented without applying any additional pulses, but rather by shifting the phases of the drive fields in the next resonant control pulses. In our experiment, these phase shifts are applied to the tomography pulses. We note that a decomposition of qutrit gates based on a single pulse with detuned tones was proposed in Ref.~\cite{ShlyakhovQuantumMetrology2018}. Simultaneous driving of three-state subspaces in superconducting devices was applied to quantum emulation~\cite{tanTopologicalMaxwellMetal2018} and adiabatic transfer protocols~\cite{vepsalainen2020simulating}.

\begin{figure}[hbt!]
\centering
\includegraphics[width=3.4in]{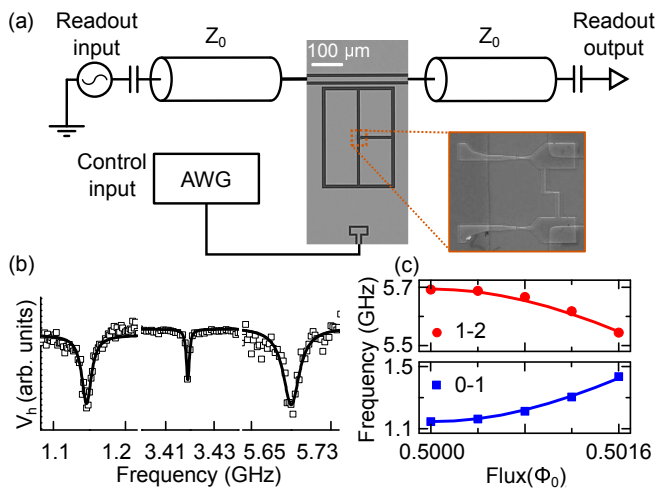}
\caption{\label{fig:fig1}(Color online) (a) Schematic and microscope images of the device used in the experiment. The device is coupled to a coplanar waveguide resonator with characteristic impedance Z$_0$ for readout and a drive pad for control with pulses generated by an arbitrary waveform generator. The inset shows the zoomed-in view of the device junctions and the loop. (b) Readout homodyne voltage $V_\t{h}$ versus the frequency of the spectroscopy tone applied to the qutrit at the flux symmetry point, showing the 0-1, two-photon 0-2, and 1-2 transitions, respectively.  The solid lines represent Lorentzian fits. (c) Peak positions for the 0-1 (blue squares) and 1-2 (red dots) transitions versus applied magnetic flux. The solid lines represent the fits with the circuit model based on the calculated qutrit energies. }
\end{figure}

The device used in our experiments, shown in Fig.~\ref{fig:fig1}(a), is formed of a superconducting loop with three Josephson junctions and three large capacitor pads. The device is capacitively coupled to a co-planar waveguide half-wavelength resonator for dispersive readout and to a transmission line terminated by a capacitor pad for control~\cite{yurtalan_anharmoniccsfq_2020}. Similar devices, based on three Josephson junction loops with capacitive shunts, were employed as qubits encoded in the lowest two energy levels~\cite{you_low-decoherence_2007,steffen_high-coherence_2010,yan_flux_2016}. We employ control pulses generated using direct synthesis by a fast arbitrary waveform generator, model Tektronix TEK70001A, with a sampling rate of 50 GS/s. The pulses consist of single- and multi-tone signals with an envelope that has cosine shape rise and fall parts and a flat top. The device loop is biased with a magnetic flux generated by external superconducting coils. The device is placed inside a sample holder at the mixing chamber of a dilution refrigerator at 27 mK. All transmission lines contain attenuators, microwave low-pass filters and infrared filters.

We first perform spectroscopy measurements to characterize the qutrit. Figure~\ref{fig:fig1}(b) shows the results of spectroscopy of the device at the flux symmetry point $\Phi = \Phi_0 / 2$. We identify the 0-1 and 1-2 transitions at frequencies $\omega_{01}=2\pi\times 1.146$ GHz and $\omega_{12}=2\pi\times 5.693$ GHz, respectively. The 0-2 transition, which is forbidden at the symmetry point, is visible as a two-photon excitation process at frequency $\omega_\t{02}^\t{2p}  = 2\pi\times 3.420$ GHz. The 0-1 and 1-2 transition peak widths change linearly with driving amplitude whereas the 0-2 peak width has a quadratic dependence on the applied driving amplitude, as expected for one and two photon processes respectively. Figure~\ref{fig:fig1}(c) shows the flux dependence of the 0-1 and 1-2 transition frequencies. The spectroscopy data is in excellent agreement with calculations based on a circuit model (see Supplemental Material) where the capacitance matrix is extracted from electromagnetic simulations and the junction critical currents are slightly adjusted with respect to nominal microfabrication values to fit the data. 

Next, we discuss the qutrit thermal state populations and implementation of qutrit state tomography. The thermal state occupation probabilities $P_\t{th0}$ and $P_\t{th1}$ for states 0 and 1 are determined by comparing the amplitudes of the Rabi oscillations in the 0-1 transition with two different initial states: the qutrit thermal state with populations of states 1 and 2 swapped, and the thermal state with the populations of states 0 and 2 swapped~\cite{yurtalan_anharmoniccsfq_2020}. Here we ignore the thermal state population $P_\t{th2}$ of state 2, which gives a good approximation for $P_\t{th0}$ and $P_\t{th1}$ given that the temperature is low relative to the frequency of the 1-2 transition. The average thermal state population in the ground state is $P_\t{th0}=0.74 \pm 0.01$. State measurement is done using dispersive readout~\cite{wallraff_approaching_2005,bianchetti_control_2010}. The average homodyne voltage $V_\t{h} = P_0 V_\t{h0} + P_1 V_\t{h1} + P_2 V_\t{h2}$, where $V_\t{h0}$, $V_\t{h1}$, and $V_\t{h2}$ are average voltages corresponding to the three qutrit states and $P_0$, $P_1$, and $P_2$ are the occupation probabilities of the qutrit states immediately prior to the measurement. The experimental procedure used to determine the voltage levels $V_\t{h0}$, $V_\t{h1}$, and $V_\t{h2}$ is given in the Supplemental Material. 

\begin{table}[h]
	\caption{Set of pulses used in state preparation and state tomography experiments.}
	\centering
	\begin{tabular*}{3.4in}{ c c @{\extracolsep{\fill} } c c c }
		\hline
		\hline
		State & Rotations &&Tomography & Rotations \\
		prep.& & & pulses& \\
		
		\cline{1-2}
		\cline{4-5}
	   p$_0$&I& &u$_0$  & $R_{x}^{01}(\pi)$  \\
	   p$_1$&$R_{x}^{01}(\pi)$& &u$_1$  & $R_{x}^{01}(\pi/2)$  \\
	   p$_2$&$R_{x}^{12}(\pi) R_{x}^{01}(\pi)$ & &u$_2$  & $R_{y}^{01}(\pi/2)$  \\
	   p$_3$&$R_{x}^{01}(\pi/2)$&& u$_3$  & I  \\
	   p$_4$&$R_{y}^{01}(\pi/2)$& &u$_4$  & $R_{x}^{12}(\pi/2)R_{x}^{01}(\pi)  $  \\
	   p$_5$&$R_{x}^{12}(\pi/2) R_{x}^{01}(\pi)$&& u$_5$  & $R_{y}^{12}(\pi/2)R_{x}^{01}(\pi)  $   \\
	   p$_6$&$R_{y}^{12}(\pi/2) R_{x}^{01}(\pi)$& &u$_6$  & $R_{x}^{01}(\pi)  R_{x}^{12}(\pi/2)  R_{x}^{01}(\pi)$  \\
	   p$_7$&$R_{x}^{12}(\pi) R_{x}^{01}(\pi/2)$&& u$_7$  & $R_{x}^{01}(\pi)  R_{y}^{12}(\pi/2)  R_{x}^{01}(\pi)$   \\
	   p$_8$&$R_{x}^{12}(\pi) R_{y}^{01}(\pi/2)$& &u$_8$  & $R_{x}^{01}(\pi)  R_{x}^{12}(\pi)  R_{x}^{01}(\pi)$   \\
 \hline \hline
	\end{tabular*}
	
	\label{tbl:tbl1}
\end{table}
To reconstruct the density matrix $\rho$ of the qutrit state, we use a quantum state tomography procedure in which the homodyne voltage is measured following the application of each pulse from a set of nine tomography pulses (see Table~\ref{tbl:tbl1}). The tomography pulses are designed to optimize the readout by utilizing the large difference between $V_\t{h1}$ and $V_\t{h0}$; this set of pulses is different from those used in the experiment of Bianchetti \emph{et al.}~\cite{bianchetti_control_2010}, where contrast is maximum between states 1 and 2. The tomography pulses consist of combinations of rotations, denoted by $R_\alpha^{01(12)}(\theta)$, where $\alpha=x,y$ is the rotation axis and $\theta$ is the rotation angle.
\begin{figure}[h]
\centering
\includegraphics[width=3.4in]{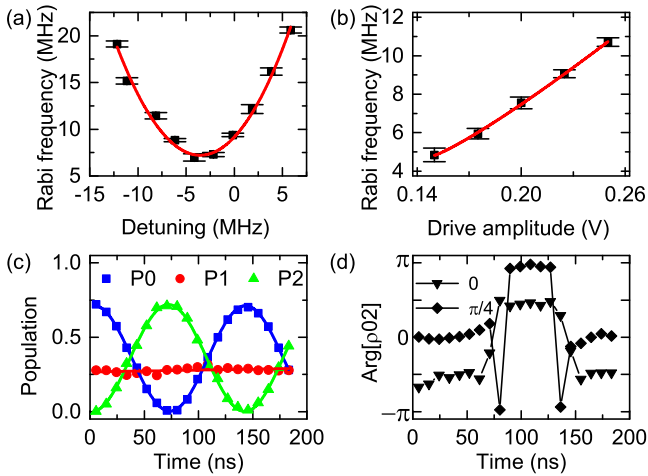}
\caption{\label{fig:fig2} (Color online) Two-photon driving experiments. (a) Frequency of the Rabi oscillations versus the detuning  between the driving frequency and the two-photon resonance frequency $(\omega_{01}+\omega_{12})/2$ for a given drive amplitude. The line is a quadratic fit. (b) Two-photon resonant-driving Rabi frequency versus drive amplitude. The line is a quadratic fit. (c,d) Results of the tomography experiments. Panel (c) shows the population of states 0 (blue squares), 1 (red disks), and 2 (green triangles), versus Rabi pulse duration, with an oscillation frequency of $2\pi\times 7.2$ MHz. Panel (d) shows the phase of the 0-2 component of the density matrix for two values of the phase of the driving field: 0 (triangles) and $\pi/4$ (rhombuses).}
\end{figure}

The implementation of the Walsh-Hadamard gate discussed above requires the simultaneous driving of all three qutrit transitions such that the effective Hamiltonian in the rotating frame has non-zero values for all off-diagonal matrix elements. While the 0-1 and 1-2 transitions are allowed and can be implemented by standard resonant driving, the 0-2 transition is forbidden at the symmetry point, and is therefore implemented as a two-photon process. Figure~\ref{fig:fig2}(a) shows the Rabi frequency of the two-photon oscillations between states 0 and 2 versus the detuning $\delta_{02}^\t{2p} = (\omega_{01}+\omega_{12})/2 - \omega_\t{d,02}^\t{2p}$, where $\omega_\t{d,02}^\t{2p}$ is the driving frequency. The resonant Rabi oscillations have a  minimum frequency of $2\pi\times 7.2$ MHz at a detuning $\delta_{02}^\t{2p}=-2\pi\times 3.5$ MHz. The fact that the resonance for the Rabi oscillations occurs at a non-vanishing detuning is due to fact that the strong field of 0-2 two-photon drive induces non-negligible ac-Stark and Bloch-Siegert shifts on the energy levels. On resonance, the frequency of the Rabi oscillations depends quadratically on the drive amplitude (Fig.~\ref{fig:fig2}(b)), characteristic of a two-photon process. We performed tomography experiments for various durations of the 0-2 Rabi pulse, as shown in Fig.~\ref{fig:fig2}(c,d). The population oscillates between states 0 and 2, while the population of state 1 remains relatively constant at the level of the thermal state. The argument of the $\bra{0} \rho \ket{2}$ element of the density matrix $\rho$, shown in Fig.~\ref{fig:fig2}(d), is constant during each half-oscillation period, indicating rotation around a constant axis in the $\{ \ket{0}, \ket{2} \}$ subspace. The two data sets in Fig.~\ref{fig:fig2}(d) correspond to two values of the phase of the two-photon driving tone, different by $\pi/4$. The phase of $\bra{0} \rho \ket{2}$ changes by twice the driving tone phase, which is another indication that the transition is a two-photon process. The results of the two-photon driving tomography experiment agree with numerical simulations based on the multi-level Hamiltonian, with the amplitude of the driving voltage at the qutrit being the only adjustable parameter. We note that in the numerical simulations we only obtained good agreement when taking into account at least the lowest seven energy levels, underscoring the importance of ac-Stark and Bloch-Siegert shifts in the experiment.

We next present the characterization of the Walsh-Hadamard gate. We use a decomposition where the off-diagonal generator has coefficients $m_{01} =0.3491+0.6046i $, $m_{12} = -0.6981 $, and $m_{02} = 0.3491+0.6046i$ and the diagonal generator has elements $\phi_0 = 6.1086$, $\phi_1 =4.0143 $, and $\phi_2 = 4.0143$. As mentioned above, the off-diagonal Hamiltonian that generates $U_\t{o}$ is obtained by the simultaneous driving of the transitions 0-1, 1-2, and 0-2, with the latter being a two-photon process. The Rabi frequencies are $\tilde\Omega_{01} = \tilde\Omega_{12}=\tilde\Omega_{02}=2 \pi \times 7.2$~MHz. The Rabi frequencies are all equal because $|m_{01}|=|m_{02}|=|m_{12}|$, and they are maximized within the available voltage range of our pulse generation setup. We note that the control signal amplitude is not limited by the qutrit properties and larger driving amplitudes could in principle be reached by increasing signal transmission and the coupling to the capacitive driving line. During the control pulse, the drive frequencies are dynamically adjusted to match the transition frequency with ac-Stark and Bloch-Siegert shifts included, with the latter dependent on the driving amplitude (see Supplemental Material).  

The frequency shift of the 0-1 transition, which includes contributions from the ac-Stark and Bloch-Siegert shifts, is experimentally determined as follows. A driving field is applied with a detuning of $2\pi\times50$ MHz from the 0-2 two-photon resonance. This detuning is chosen to be much larger than the strength of the two-photon process (when applied on resonance). Therefore, transitions between the 0 and 2 states are negligible, while the ac-Stark and Bloch-Siegert shifts are almost the same as those induced in the case of resonant driving. The shifts can then be measured directly by applying a microwave pulse to drive the 0-1 transition and measuring the induced Rabi oscillations. The driving frequency at which the Rabi oscillations have their minimum frequency is a direct measure of the resonance frequency, which includes the shifts caused by the strong 0-2 drive. The experimentally determined total shift of $2\pi\times9.4$ MHz is in agreement with that obtained from numerical simulations.

\begin{figure}[hbt!]
\centering
\includegraphics[width=3.4in]{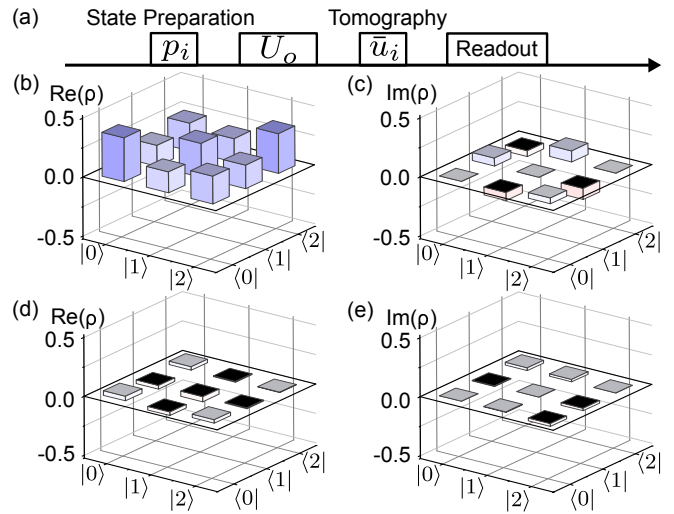}
\caption{\label{fig:fig3} (a) Pulse sequence used in the experiments. (b) Real and (c) imaginary parts of the reconstructed density matrix after the gate is applied on the thermalized state shows a 99.8\% gate fidelity. The differences between the reconstructed and  the expected density matrices are shown for the real (d) and imaginary (e) parts.}
\end{figure}

The diagonal Hamiltonian that generates $U_\t{d}$ is effectively embedded in the tomography pulses by shifting the phases of the driving fields for the latter. The new tomography analyzer pulses are given by $\bar{u}_i= U_\t{d}^\dag u_iU_\t{d}$. The pulse sequence is shown in Fig.~\ref{fig:fig3}(a). The off-diagonal Hamiltonian is applied on the prepared states $p_i$ followed by the phase-shifted tomography pulses $\bar{u}_i$ to complete the Walsh-Hadamard gate and to reconstruct the density matrix $\rho$ of the state after the gate. 

The gate fidelity is measured with respect to the ideal evolution of the prepared state under the gate Hamiltonian in Eq.~(\ref{eq:cgate}). Figure~\ref{fig:fig3} shows the real (b) and imaginary (c) elements of $\rho$  where the gate is applied on the qutrit thermal state. The state fidelity after the gate in this case is 99.8\%. The difference between the reconstructed and the expected density matrices are shown in Fig.~\ref{fig:fig3}(d) and (e) for the real and imaginary parts, respectively. Table~\ref{tbl:tbl2} shows the fidelity values of the gate applied on 9 different prepared states. On average over the 9 preparations, the state fidelity at the end of the sequence is observed to be $99.2\pm0.1$\%. We estimate the error in the evolution by applying a maximum likelihood estimation procedure (see Supplemental Material) together with the  errors  associated with the state homodyne voltage levels. We also utilize quantum process tomography to quantify the process fidelity (see Supplemental Material). The process fidelity of the Walsh-Hadamard gate is found to be 97.3\%.

\begin{table}[hbt!]
	\caption{Set of pulses used to prepare different states to which the Walsh-Hadamard gate is applied and measured fidelities.}
	\centering
	\begin{tabular*}{3.4in}{c @{\extracolsep{\fill}} c c c }
		\hline
		\hline
		State prep.($p_i$)  & Fidelity & State prep.($p_i$) & Fidelity\\ 
		\cline{1-2}
		\cline{3-4}
	    I                                      & 99.8\% & $R_{x}^{12}(\pi/2)$ $R_{x}^{01}(\pi)$ & 98.7\% \\
	    $R_{x}^{01}(\pi)$                      & 99.8\% & $R_{y}^{12}(\pi/2)$ $R_{x}^{01}(\pi)$  & 98.3\% \\
	    $R_{x}^{12}(\pi)$ $R_{x}^{01}(\pi)$    & 98.8\% & $R_{x}^{12}(\pi)$ $R_{x}^{01}(\pi/2)$  & 99.6\% \\
	    $R_{x}^{01}(\pi/2)$                    & 99.8\% & $R_{x}^{12}(\pi)$ $R_{y}^{01}(\pi/2)$  & 98.5\% \\
	    $R_{y}^{01}(\pi/2)$                    & 99.6\% &  \\
 \hline \hline
	\end{tabular*}
	\label{tbl:tbl2}

\end{table}

\begin{figure}[h]
\centering
\includegraphics[width=3.4in]{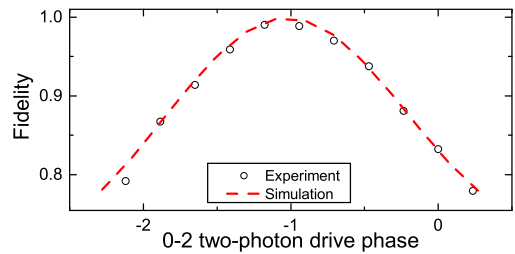}
\caption{\label{fig:fig4} (Color online) Experimental (dots) and simulated (dashed curve) fidelity of the state obtained by applying the control pulse to the thermal state versus the relative phase of the 0-2 two-photon drive.
}
\end{figure}

We next discuss numerical simulations of the evolution of the set of states obtained by applying the pulses listed in Table~\ref{tbl:tbl2} to the qutrit steady state under the applied control pulse. The simulations are based on the system Hamiltonian with driving including the lowest seven energy levels of the system. The transition strengths are determined numerically with  the circuit model including the complete capacitance matrix (see Supplemental Material). The simulated fidelity of the off-diagonally generated part of the gate is 99.8\% (99.7\%) without (with) decoherence (see Supplemental Material). To verify the role of the relative phase of the driving tones, we performed an experiment where the phase of the 0-2 two-photon driving tone was changed around its nominal value, based on the decomposition of the gate.  Figure~\ref{fig:fig4} shows the measured average fidelity of states after preparation by the pulses given in Table~\ref{tbl:tbl2} and application of the Walsh-Hadamard gate versus the 0-2 two-photon drive phase. The data are in excellent agreement with the average fidelity values obtained from numerical simulations (dashed line in Fig.~\ref{fig:fig4}). We note that while the numerical simulations require the inclusion of several higher levels to accurately reproduce the experimentally observed driving strength and level shifts, the experimental implementation of the gate relies only on experimentally determined driving strengths and level shifts.

The results described above were obtained using the thermal state as the initial state. While this procedure did not pose a limitation for characterizing the qutrit gate, it is relevant to consider the possibility to prepare high purity initial states. To date, several initialization and state reset protocols have been demonstrated with superconducting qubits (see \emph{e.g.}~\cite{geerlingsDemonstrating2012,ValenzuelaMicrowave2006}). In this work, we utilize a cooling technique based on a Raman process that effectively brings the qutrit 0-1 transition into resonance with the readout resonator. The fact that the resonator has a high resonance frequency and hence a small thermal excitation probability creates an effectively cold environment for the qutrit. By controlling the rate of transferring excitations between the qutrit and the resonator, the cooled qutrit state reaches a maximum ground state probability of 0.94, a significant improvement over the probability of 0.74 in the thermal state. With the cooled state as the initial state for the Walsh-Hadamard gate, we measure an average state fidelity of 97.3\% and a quantum process fidelity of 97.3\%. While the process fidelity is identical within experimental errors for thermal state and cooling based preparation, the average state fidelity is slightly lower in the latter case. This is possibly due to spurious excitation of the resonator having a reduced effect in combination with the strong state preparation pulses used in process tomography. The details and an extended analysis of the cooling process and quantum control on reset states are beyond the scope of the present work and will be addressed in future work.

In conclusion, we have implemented a Walsh-Hadamard gate in a superconducting qutrit. The implementation of the gate relied on a two-step decomposition, which only required the application of a single microwave pulse with three tones coupling the three pairs of qutrit energy levels. The experimentally characterized state fidelity after gate control is 99.2\%, in agreement with the results of numerical simulations. We note that the ac-Stark and Bloch-Siegert shifts of the qutrit energy levels have to be accounted for in order to achieve a high fidelity gate. Our approach can be generalized to higher dimensionality qudits. This work demonstrates the potential of multi-tone multi-level control in superconducting devices and opens interesting avenues for exploration of superconducting qudits in quantum computing and other areas of quantum science.

We thank University of Waterloo Quantum Nanofab team members for assistance on the device fabrication. We acknowledge support from NSERC, Canada Foundation for Innovation, Ontario Ministry of Research and Innovation, Industry Canada, and the Canadian Microelectronics Corporation. During this work, A. L. was supported by an Early Researcher Award.

\onecolumngrid
\clearpage

\begin{center}
	\textbf{\large {Supplemental Material - Implementation of a Walsh-Hadamard gate in a superconducting qutrit}}
\end{center}

\twocolumngrid
\setcounter{equation}{0}
\setcounter{figure}{0}
\setcounter{table}{0}
\setcounter{page}{1}
\makeatletter
\renewcommand{\theequation}{S\arabic{equation}}
\renewcommand{\thefigure}{S\arabic{figure}}
\renewcommand{\thetable}{S\arabic{table}}
\renewcommand{\bibnumfmt}[1]{[S#1]}
\renewcommand{\citenumfont}[1]{S#1}

\setcounter{secnumdepth}{2}

\section{ Experimental details}
\subsection{Sample fabrication and experimental setup}
The device used in the experiments is fabricated in two steps. The first step uses standard optical lithography and a lift-off process for the aluminum coplanar waveguide (CPW) structures and capacitor pads. In the second step, the Josephson junctions are patterned with e-beam lithography and  then shadow evaporated on the capacitor pads, forming the flux loop. The device is placed in a microwave package and mounted in a dilution refrigerator operating at 27 mK. The sample is shielded with  high magnetic-permeability shields and all the coaxial cables for readout and control are filtered and attenuated  at different stages of the fridge. The microwave control pulses are directly generated  by a fast arbitrary waveform generator(Tektronix AWG70001A) with a sampling rate of 50 GS/s. 

\subsection{Steady state populations and homodyne voltages}
The probabilities  $P_\t{th0}$, $P_\t{th1}$, and $P_\t{th2}$ for states 0, 1, and 2 in the steady state are determined as follows. Rabi oscillations on the 0-1 transition are performed for two different prepared states: i) the steady state, obtained by waiting a sufficiently long time after the previous run of the experiment, followed by a swap of states 1 and 2,  and ii) the steady state followed by a swap of populations on states 0 and 2~\cite{SIyurtalan_anharmoniccsfq_2020}. Assuming the population of state 2 to be negligible, the amplitudes of the Rabi oscillations in these two situations have a ratio $P_\t{th1}/P_\t{th0}$. The oscillation amplitudes of the measured voltages have the same ratio. Combined with the fact that $P_\t{th0}+P_\t{th1}=1$, this ratio can be used to determine $P_\t{th0}$ and $P_\t{th1}$. During the course of the experiments, $P_\t{th0}$ is found to be $0.74\pm0.01$ indicating an effective temperature of the qubit of 50-55 mK based on Boltzmann statistics.

To determine the signal levels $V_\t{h0}$, $V_\t{h1}$, and $V_\t{h2}$, the homodyne voltage $V_\t{h}$ is measured for three states: i) the qutrit thermal state, corresponding to $P_0 = P_\t{th0}$, $P_1 = P_\t{th1}$, and $P_2 = 0$, ii) the thermal state followed by a swap of populations on 0 and 1, corresponding to $P_0 = P_\t{th1}$, $P_1 = P_\t{th0}$, and $P_2 = 0$, and iii) the thermal state followed by a swap of populations on 0 and 1 and a swap of populations 1 and 2, corresponding to $P_0 = P_\t{th1}$, $P_1 = 0$, and $P_2 = P_\t{th0}$. 

\subsection{Coherence time characterization}\label{subsec:mlc}
In this section we discuss the decoherence model and experimental procedure used for extracting the coherence times of our qutrit.

In the theoretical model, we consider a situation where an initial state is prepared and then undergoes decoherence in the absence of driving. We first discuss decoherence due to energy relaxation and excitation processes; the additional role of pure dephasing will be discussed next. Relaxation/excitation is modeled by the following master equation for the density matrix $\rho$:
\begin{equation}
\dot{\rho} = -i[H,\rho]+\sum_{i<j} \Gamma_{ij}\mathcal{L}[\sigma_{ij}^+]\rho + \sum_{i>j} \Gamma_{ij}\mathcal{L}[\sigma_{ij}^-]\rho. 
\label{eq:mlc}
\end{equation}
Here $H$ is the Hamiltonian, which is time independent during the decoherence dynamics interval considered here. The rates $\Gamma_{ij}$ are the energy relaxation ($i>j$) and excitation ($i<j$) rates, and their indices $i$ and $j$ specify respectively the initial and final states of each transition. The Lindblad form superoperators are defined as $\mathcal{L}[\sigma]\rho = \sigma \rho \sigma^\dagger -\frac{1}{2} \sigma^\dagger \sigma \rho -\frac{1}{2} \rho  \sigma^\dagger \sigma $, where the operators $\sigma_{ij}^{+} = \ket{j} \bra{i}$ (defined for $i <j$) and $\sigma_{ij}^{-} = \ket{j} \bra{i}$ (defined for $i > j$) are the excitation and relaxation operators, respectively. Note that the different relaxation/excitation transitions are described by different terms in the master equation, which is appropriate for a qutrit with non-degenerate transition frequencies. The evolution of the density matrix due to Eq.~(\ref{eq:mlc}) can be written formally as $\rho(t) = \mathcal R (t-t_0) [\rho (t_0)]$, where $\mathcal R (t-t_0)$ is a superoperator that acts on the initial density matrix at time $t_0$ and produces the final density matrix at time $t$. The superoperator $\mathcal R (t-t_0)$ accounts for the free evolution of the qutrit states and relaxation/excitation induced mixing of the populations, including the appropriate amount of dephasing that necessarily accompanies relaxation/excitation.

Pure dephasing leads to additional suppression of the off-diagonal matrix elements. The overall evolution in this case is described by
\begin{equation}
\rho(t) = \mathcal D (t-t_0) \left[ \mathcal R (t-t_0) [\rho (t_0)] \right].
\label{eq:mlcevol}
\end{equation}
The superoperator $\mathcal D (t-t_0)$ leaves the diagonal elements unaffected and modifies the off-diagonal elements according to $\rho_{ij} \rightarrow \mathcal C_{ij} (t-t_0) \rho_{ij}$ ($i \neq j$), where $C_{ij} (\tau)$ is the coherence function for the pair of levels $(i, j)$, which depends on the fluctuations in this transition frequency~\cite{SIyurtalan_anharmoniccsfq_2020}. We note that we did not describe pure dephasing by an additional term in the Lindblad form master equation, because this approach would only be rigorously justifiable for noise with short correlation time. If this were the case, it would lead to exponentially decaying coherence functions $C_{ij} (\tau)$. With 1/f noise and low frequency noise in general, the coherence functions $C_{ij} (\tau)$ are not exponentially decaying functions. Indeed, we find that Gaussian decay functions are a better fit for our data, which is consistent with coherence functions measured in experiments on two-level systems (qubits) in the presence of 1/f noise.

The experiments are conducted as follows. First the decay rate $\Gamma_{01} + \Gamma_{10}$ is measured by preparing the qutrit in the first excited state and measuring the population decay rate. The ratio between the rates $\Gamma_{01}$  and $\Gamma_{10}$ is next determined based on the population measurement described in the previous subsection. This procedure yields the rates $\Gamma_{01} = 5.5$ kHz and $\Gamma_{10} =  16.2$ kHz. Next, the remaining rates are measured by preparing the qutrit in the second excited state by applying a $\pi_{x}^{01}$ and a $\pi_{x}^{12}$ pulse  and measuring the state populations versus time. We fit the populations with a model following Zizak \emph{et.al}~\cite{zizak_rate_1980} and we constrain the fit by assuming that the ratio of the relaxation and excitation rates are set to Boltzmann factors with the effective temperature of the device. We find  $\Gamma_{20} = 21.6$ kHz and $\Gamma_{21} = 314.5$ kHz 

Next the pure dephasing rates are measured as follows. First, we prepare a superposition of the qutrit states i and j in a Ramsey-type experiment, and we measure the decay of the Ramsey oscillations and fit these oscillations with the model described in Eq.~(\ref{eq:mlcevol}). By accounting for the energy relaxation and excitation rates and assuming Gaussian decay for the coherence functions, the 1/e decay times of the coherence functions give the pure dephasing rates $\Gamma_{01}^\t{R}= 204.1$ kHz, $\Gamma_{12}^\t{R}= 238.1$ kHz, and $ \Gamma_{02}^\t{R}= 181.8$ kHz, for 0-1, 1-2, and 0-2 superpositions, respectively.


\section{Superconducting Device Model}

\begin{figure}[h]
	\centering
	\includegraphics[width=2in]{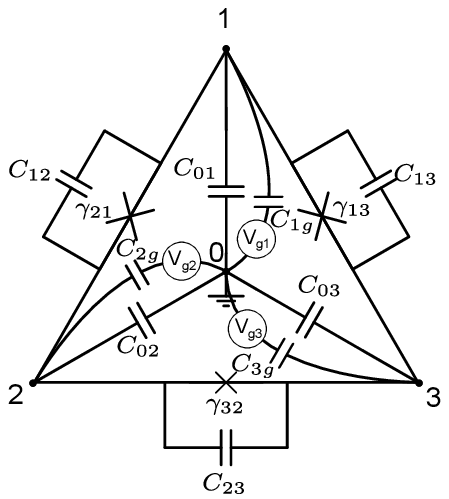}
	\caption{\label{fig:figs1} Circuit model of the capacitively shunted flux qubit.}
\end{figure}

The circuit model for the capacitively shunted qubit is shown in Fig.~\ref{fig:figs1}. The nodes 0, 1, 2, and 3 represent the CPW ground plane and three shunt capacitors pads, respectively. The resonator center electrode and drive electrode are modeled by the gate capacitors $\mathbf{C}_\t{b}=(C_\t{1b},C_\t{2b},C_\t{3b})$ and $\mathbf{C}_\t{d}=(C_\t{1d},C_\t{2d},C_\t{3d})$, respectively with corresponding bias voltages $V_\t{b}$ and $V_\t{d}$. Trapped charges can be modelled by gate capacitors and gate voltages.  The Josephson junctions, indicated by crosses, are placed between nodes and have superconducting phase differences $\gamma_{21}$, $\gamma_{31}$, and $\gamma_{23}$. The circuit Hamiltonian is given by
\begin{equation}
H=T_\t{H}+U_\t{H},\label{eq:ham}  
\end{equation}
where the kinetic term is expressed as 
\begin{eqnarray}
T_\t{H}&=&\frac{1}{2\varphi_0} \big [ \mathbf{p}  - \varphi_0 \mathbf{D} ( {V}_\t{b}\mathbf{C}_\t{b} + {V}_\t{d}\mathbf{C}_\t{d} + \mathbf{Q}_\t{g}  )  \big ] \nonumber \\ && \mathbf{C}^{-1} \big [ \mathbf{p}  - \varphi_0 \mathbf{D} ( {V}_\t{b}\mathbf{C}_\t{b} + {V}_\t{d}\mathbf{C}_\t{d} + \mathbf{Q}_\t{g}  )   \big ]^T,\label{eq:hamt}
\end{eqnarray}
and the potential term is given by
\begin{eqnarray}
U_\t{H}&=&-\varphi_0 I_\t{c}[  \cos(\gamma_{21}) + \cos(\gamma_{31}) \nonumber \\  &&+\alpha \cos(\gamma_{21}-\gamma_{31}+2\pi f) ],
\end{eqnarray}
where $\varphi_0=\Phi_0/2\pi$ is the reduced flux quantum and $f=\Phi /\Phi_0$, with $\Phi$ the external magnetic flux.  $\mathbf{p}=(p_{\gamma_{21}},p_{\gamma_{31}},p_{\gamma_{01}})$ is the momentum vector with components given by the variables conjugate to the corresponding phases. The junction critical current is represented by $I_\t{c}$ for the two equal-area junctions and it is smaller by a factor $\alpha=0.61$ for the smaller-area junction. The capacitance matrix $\mathbf{C}$ is given by
\begin{equation}
\mathbf{C} =
\begin{pmatrix}
\begin{smallmatrix}  
\substack{C_{12}+C_{23}C_\t{02}\\+C_\t{2b}+C_\t{2d} }  &-C_{23} 			  & \substack{C_{02}+C_\t{2b}+C_\t{2d}}\\ \\
-C_{23}		           & \substack{C_{13}+C_{23}+C_\t{03}\\+C_\t{3b}+C_\t{3d}}  & \substack{C_{03}+ C_\t{3b} +C_\t{3d}}\\ \\
\substack{C_{02}+C_\t{2b}+C_\t{2d}}    & \substack{C_{03}+C_\t{3b} +C_\t{3d}}	  & \substack{C_{01}+C_{02}+C_{01} \\ +C_\t{1b}+C_\t{2b}+C_\t{3b}  \\  +C_\t{1d}+C_\t{2d}+C_\t{3d}} 
\end{smallmatrix}

\end{pmatrix}
\label{eq:cmatrix}
\end{equation}
and the matrix $\t{D}$ is given by
\begin{equation}
\mathbf{D}=
\begin{pmatrix}
0 &-1 & 0 \\
0 & 0 &-1 \\
-1&-1 &-1
\end{pmatrix}.   
\end{equation}

\section{Decomposition of the Walsh-Hadamard gate into two unitary operations, generated by a diagonal matrix and an off-diagonal matrix respectively}
As mentioned in the main text, we seek a decomposition of the Walsh-Hadamard gate of the form
\begin{equation}
U_\t{WH} = U_\t{d} U_\t{o},
\label{Eq:UdUodDecomposition}
\end{equation}
where $U_\t{d}  =  \exp(-iG_\t{d})$, $U_\t{o}  = \exp(-iG_\t{o})$, and the generators $G_\t{d}$ and $G_\t{o}$ have the form 
\begin{eqnarray}
G_\t{d} & = & \left(
\begin{array}{ccc}
\phi_0 & 0 & 0 \\
0 & \phi_1 & 0 \\
0 & 0 & \phi_2
\end{array}
\right), \nn \\
G_\t{o} & = & \left(
\begin{array}{ccc}
0 & m_{01} & m_{02} \\
m_{01}^* & 0 & m_{12} \\
m_{02}^* & m_{12}^* & 0
\end{array}
\right).
\label{Eq:UdUodDecompositionDetails}
\end{eqnarray}
We perform a numerical search where we sweep the phases $\phi_0,\phi_1,\phi_2$ between 0 and $2\pi$. For each set of phases, we use Eq.~(\ref{Eq:UdUodDecomposition}) to evaluate $U_\t{o}$ and then use Eq.~(\ref{Eq:UdUodDecompositionDetails}) to evaluate the corresponding generator $G_\t{o}$. If the diagonal matrix elements of $G_\t{o}$ vanish, we identify a valid decomposition of the target gate. For $U_\t{WH}$ we find five decompositions listed in Table~\ref{tbl:tblS1}. 
\begin{table*}[t]
	\caption{Numerically determined matrix elements of Walsh-Hadamard gate generators when the gate is decomposed into two unitary operators as explained in the text.}
	\label{tbl:tblS1}
	\centering
	\begin{tabular*}{7in}{@{\extracolsep{\fill} } c c c c c c c }
		\hline
		\hline
		Decomposition & $m_{01}$ & $m_{12}$ & $m_{02}$ &$\phi_0$&$\phi_1$&$\phi_2$ \\
		\hline
		1&$-0.9672-0.2365i$&$1.9345$ &$-0.9672-0.2365i$& $0.8434$ & $0.3637$ & $0.3637$ \\
		2&$-0.6982-1.2092i$&$1.3962$ &$-0.6981-1.2092i$& $1.9199$ & $6.1087$ & $6.1086$ \\
		3&$-0.9672-1.6753i$&$0.6885+0.7194i$ &$0.2788-0.9559i$& $2.4581$ & $0.3637$ & $5.0322$ \\
		4&$ 0.2788-0.9559i$&$0.6885-0.7194i$ &$-0.9672-1.6753i$& $2.4581$ & $5.0322$ & $0.3637$ \\
		5&$0.3491+0.6046i$&$-0.6981$ &$0.3491+0.6046i$& $6.1086$ & $4.0143$ & $4.0143$ \\
		\hline 
		\hline
	\end{tabular*}
\end{table*}
Any of these five decompositions can be used to construct the Walsh-Hadamard gate in the experiment. We note, however, that the Rabi frequencies of the driven gate dynamics will be proportional to the matrix elements $|m_{ij}|$. More specifically, assuming a square pulse envelope and weak driving, the product of the Rabi frequency and the gate pulse duration should be equal to $2|m_{ij}|$. The driving field amplitudes will in turn depend on $|m_{ij}|$, with a linear relation for the 0-1 and 1-2 transitions and a quadratic relation for the 0-2 transition, because we drive the 0-2 transition using a two-photon process. Because the two-photon transition requires stronger driving fields than those required for the other two transitions, and we would like to obtain the fastest gate for a given drive power, we choose the decomposition that has the smallest value of $|m_{02}|$, which is the fifth decomposition in Table~\ref{tbl:tblS1}.

\section{AC-Stark and Bloch-Siegert Shift Modeling}
Because we use a two-photon process to drive the 0-2 transition, we apply a relatively strong drive field at the frequency $\omega_{02}/2$, which causes non-negligible ac-Stark and Bloch-Siegert shifts in the 0-1 and 1-2 transition frequencies. These shifts must be taken into account when choosing the pulse parameters needed to obtain a high-fidelity implementation of the gate. In this Section we discuss how these shifts are calculated.

If we drive the 0-2 transition resonantly using a two-photon process with a given drive power and driving field frequency $\omega_\t{d}=\omega_{02}/2$, the resulting Rabi oscillation frequency is given by
\begin{equation}
\Omega_\t{Rabi,tp} = \left|\frac{\Omega_{01}\Omega_{12}}{2(\omega_\t{d}-\omega_{01})}\right|,
\label{eq:TwoPhotonRabiFreq}
\end{equation}
where $|\Omega_{01}|$ is the Rabi frequency of 0-1 oscillations if the same drive power is applied resonantly with the 0-1 transition, $|\Omega_{12}|$ is the Rabi frequency of 1-2 oscillations if the same drive power is applied resonantly with the 1-2 transition, and $\omega_{01}$ is the resonance frequency of the 0-1 transition (including any frequency shifts in the experiment). 

The same drive field that is used to drive the 0-2 two-photon transition can be seen as an off-resonant drive field for the 0-1 and 1-2 transitions. As a result, all of these energy levels will experience ac-Stark shifts. If we consider only the 0-1 transition, and considering that $\omega_\t{d}>\omega_{01}$, the ac-Stark shift will bring levels 0 and 1 closer to each other (i.e.~a negative ac-Stark shift) by
\begin{equation}
\delta\omega_{01} = - \frac{|\Omega_{01}|^2}{2(\omega_\t{d}-\omega_{01})}.
\label{eq:01acStarkShift}
\end{equation}
A similar expression can be obtained for the ac-Stark shift of the 1-2 transition:
\begin{equation}
\delta\omega_{12} = \frac{|\Omega_{12}|^2}{2(\omega_\t{d}-\omega_{01})}.
\label{eq:12acStarkShift}
\end{equation}
Since $|\Omega_{01}|$ and $|\Omega_{12}|$ are comparable to each other, Eqs.~(\ref{eq:TwoPhotonRabiFreq}-\ref{eq:12acStarkShift}) indicate that the two ac-Stark shifts are comparable to the two-photon Rabi frequency. Furthermore, in our implementation of the gate all transitions are driven simultaneously with comparable Rabi frequencies for all pairs of transitions. As a result, the fields used to resonantly drive the 0-1 and 1-2 transitions will have Rabi frequencies that are comparable to the Rabi frequency of the two-photon transition and hence comparable to the 0-1 and 1-2 ac-Stark shifts. As is well known from the basic theory of Rabi oscillations, if we have a detuning from exact resonance that is comparable to the resonant Rabi frequency, the Rabi oscillation amplitude will be reduced significantly from its resonant value, which is one (i.e.~full population transfer in the energy level pair that is being driven). For this reason, the ac-Stark shifts must be taken into account and the drive field frequencies must all be shifted to correct for the ac-Stark shifts. Since the detuning $\omega_\t{d}-\omega_{01}$ is comparable to $\omega_{01}$ in our experiment, the Bloch-Siegert shift, which is given by formulas similar to those in Eqs.~(\ref{eq:01acStarkShift},\ref{eq:12acStarkShift}) but with positive signs in the denominator, this shift will not be much smaller than the ac-Stark shift, and it will also be non-negligible. 

So far in this section we have used qualitative arguments in which we have assumed that the device under consideration contains only three quantum states and we have used the expressions for the frequency shifts that are suitable for an isolated pair of energy levels. Our device, however, contains higher energy levels that contribute additional shifts that are comparable to those given above and can therefore lead to significant deviations from the three-level approximation. For example, because $|\Omega_{12}|>|\Omega_{01}|$, the three-level approximation would suggest that the sum of the ac-Stark shifts should induce a net increase in the 0-2 transition frequency, but in the experiment we find that the 02 transition frequency is shifted in the opposite direction under the influence of the strong two-photon drive field. By gradually increasing the number of energy levels in the theoretical model, we find that we must keep at least seven energy levels to obtain good agreement between the theoretical model and experimental data.

The dressed-state picture is well-suited for the calculation of the frequency shifts. In the three-level approximation, the dressed-state Hamiltonian can be expressed as
\begin{widetext}
	\begin{equation}
	H_\t{ds} = \left( \begin{array}{ccccccccccc}
	\ddots & & & & & & & & & & \\
	& 0 & 0 & 0 & 0 & \frac{\Omega_{01}}{2} & \frac{\Omega_{02}}{2} & 0 & 0 & 0 & \\
	& 0 & \omega_{01} & 0 & \frac{\Omega_{01}^*}{2} & 0 & \frac{\Omega_{12}}{2} & 0 & 0 & 0 & \\
	& 0 & 0 & \omega_{02} & \frac{\Omega_{02}^*}{2} & \frac{\Omega_{12}^*}{2} & 0 & 0 & 0 & 0 & \\
	& 0 & \frac{\Omega_{01}}{2} & \frac{\Omega_{02}}{2} & \omega_\t{d} & 0 & 0 & 0 & \frac{\Omega_{01}}{2} & \frac{\Omega_{02}}{2} & \\
	& \frac{\Omega_{01}^*}{2} & 0 & \frac{\Omega_{12}}{2} & 0 & \omega_{01}+\omega_\t{d} & 0 & \frac{\Omega_{01}^*}{2} & 0 & \frac{\Omega_{12}}{2} & \\
	& \frac{\Omega_{02}^*}{2} & \frac{\Omega_{12}^*}{2} & 0 & 0 & 0 & \omega_{02}+\omega_\t{d} & \frac{\Omega_{02}^*}{2} & \frac{\Omega_{12}^*}{2} & 0 & \\
	& 0 & 0 & 0 & 0 & \frac{\Omega_{01}}{2} & \frac{\Omega_{02}}{2} & 2\omega_\t{d} & 0 & 0 & \\
	& 0 & 0 & 0 & \frac{\Omega_{01}^*}{2} & 0 & \frac{\Omega_{12}}{2} & 0 & \omega_{01}+2\omega_\t{d} & 0 & \\
	& 0 & 0 & 0 & \frac{\Omega_{02}^*}{2} & \frac{\Omega_{12}^*}{2} & 0 & 0 & 0 & \omega_{02}+2\omega_\t{d} & \\
	& & & & & & & & & & \ddots
	\end{array}
	\right).
	\label{eq:3LevelDressedStateHamiltonian}
	\end{equation}
\end{widetext}
The dressed-state Hamiltonian for the seven-level approximation of the superconducting circuit can be constructed straightforwardly in a similar manner. The energy shifts and Rabi frequencies can be obtained with high accuracy by numerically diagonalizing a relatively small (e.g.~$50 \times 50$) truncated version of the infinite-size matrix.

Let us assume that the driving fields are weak, and hence $|\Omega_{ij}|$ are all small compared to the detunings in the system. In this case we can use a perturbation-theory approach to obtain analytic expressions for the frequency shifts. These results turn out to give accurate results for the driving strengths that we use in the experiment. Since in this section we want to calculate the frequency shifts, we can also assume for this calculation that $\omega_\t{d}$ is not tuned exactly to the two-photon resonance. We can now calculate the frequency shifts by considering the different off-diagonal contributions one by one.

If we take Eq.~(\ref{eq:3LevelDressedStateHamiltonian}) and ignore $\Omega_{02}$ and $\Omega_{12}$, we obtain the Hamiltonian
\begin{widetext}
	\begin{equation}
	H_\t{ds} = \left( \begin{array}{ccccccccccc}
	\ddots & & & & & & & & & & \\
	& 0 & 0 & 0 & 0 & \frac{\Omega_{01}}{2} & 0 & 0 & 0 & 0 & \\
	& 0 & \omega_{01} & 0 & \frac{\Omega_{01}^*}{2} & 0 & 0 & 0 & 0 & 0 & \\
	& 0 & 0 & \omega_{02} & 0 & 0 & 0 & 0 & 0 & 0 & \\
	& 0 & \frac{\Omega_{01}}{2} & 0 & \omega_\t{d} & 0 & 0 & 0 & \frac{\Omega_{01}}{2} & 0 & \\
	& \frac{\Omega_{01}^*}{2} & 0 & 0 & 0 & \omega_{01}+\omega_\t{d} & 0 & \frac{\Omega_{01}^*}{2} & 0 & 0 & \\
	& 0 & 0 & 0 & 0 & 0 & \omega_{02}+\omega_\t{d} & 0 & 0 & 0 & \\
	& 0 & 0 & 0 & 0 & \frac{\Omega_{01}}{2} & 0 & 2\omega_\t{d} & 0 & 0 & \\
	& 0 & 0 & 0 & \frac{\Omega_{01}^*}{2} & 0 & 0 & 0 & \omega_{01}+2\omega_\t{d} & 0 & \\
	& 0 & 0 & 0 & 0 & 0 & 0 & 0 & 0 & \omega_{02}+2\omega_\t{d} & \\
	& & & & & & & & & & \ddots
	\end{array}
	\right).
	\end{equation}
\end{widetext}
A straightforward $2 \times 2$ matrix diagonalization shows that, to second order in $|\Omega_{01}|$, the levels at energies $n\omega_\t{d}$ acquire an ac-Stark shift of
\begin{equation}
\delta\omega_0 = \frac{|\Omega_{01}|^2}{4(\omega_\t{d}-\omega_{01})},
\end{equation}
and the levels at energies $\omega_{01}+n\omega_\t{d}$ acquire an ac-Stark shift of
\begin{equation}
\delta\omega_1 = - \frac{|\Omega_{01}|^2}{4(\omega_\t{d}-\omega_{01})},
\end{equation}
while the levels at energies $\omega_{02}+n\omega_\t{d}$ are not affected by the off-diagonal matrix elements and are therefore not shifted. In other words, the ac-Stark shift [Eq.~(\ref{eq:01acStarkShift})] is the result of two equal contributions from shifts in the energy levels 0 and 1, while other levels are not affected by the driving of the 01 transition. This rule can be applied to all the transitions in the system to calculate the net ac-Stark shifts for all energy levels. Thus we obtain the following shifts for the lowest three energy levels in the $N=7$ level approximation:
\begin{eqnarray}
\delta_{0,\rm acS} & = & \sum_{j=1}^{N-1}\frac{|\Omega_{0j}|^2}{4(\omega_\t{d}-\omega_{0j})} \\
\delta_{1,\rm acS} & = & \frac{-|\Omega_{01}|^2}{4(\omega_\t{d}-\omega_{01})} + \sum_{j=2}^{N-1}\frac{|\Omega_{1j}|^2}{4(\omega_\t{d}-\omega_{1j})} \\
\delta_{2,\rm acS} & = & \sum_{j=0}^1\frac{-|\Omega_{j2}|^2}{4(\omega_\t{d}-\omega_{j2})} + \sum_{j=3}^{N-1}\frac{|\Omega_{2j}|^2}{4(\omega_\t{d}-\omega_{2j})}
\end{eqnarray}
As mentioned above, the two-photon drive frequency $\omega_\t{d}$ in our experiment is so far detuned from the transition frequencies (e.g.~$\omega_{01}$) that the Bloch-Siegert shift is not much smaller than the ac-Stark shift. The calculation of the Bloch-Siegert shift is similar to the ac-Stark shift calculation given above, keeping in mind that the Bloch-Siegert shift is always positive. This calculation leads to the energy level shifts:
\begin{eqnarray}
\delta_{0,\rm BS} & = & \sum_{j=1}^{N-1}\frac{-|\Omega_{0j}|^2}{4(\omega_\t{d}+\omega_{0j})} \\
\delta_{1,\rm BS} & = & \frac{|\Omega_{01}|^2}{4(\omega_\t{d}+\omega_{01})} - \sum_{j=2}^{N-1}\frac{|\Omega_{1j}|^2}{4(\omega_\t{d}+\omega_{1j})} \\
\delta_{2,\rm BS} & = & \sum_{j=0}^1\frac{|\Omega_{j2}|^2}{4(\omega_\t{d}+\omega_{j2})} - \sum_{j=3}^{N-1}\frac{|\Omega_{2j}|^2}{4(\omega_\t{d}+\omega_{2j})} 
\end{eqnarray}
The sum of the above expressions for the ac-Stark shifts and Bloch-Siegert shifts accurately describes the shifts that we observe in the experiment. For the 0-2 two-photon drive in the experiments, calculated total shifts in the 0-1 and 1-2 transition frequencies, including the first seven levels of the device, are 
\begin{eqnarray}
\delta_\t{01}^\t{ds} & = & -2\pi\times9.2\, \t{MHz} \\
\delta_\t{12}^\t{ds} & = & 2\pi\times2.7\, \t{MHz}.
\end{eqnarray}
These results are in good agreement with the experiments, given the use of perturbative expressions and the fact that the cavity is not explicitly included in this model.

\section{Experimental determination of frequency shifts}

\begin{figure}[h]
	\centering
	\includegraphics[width=3.4in]{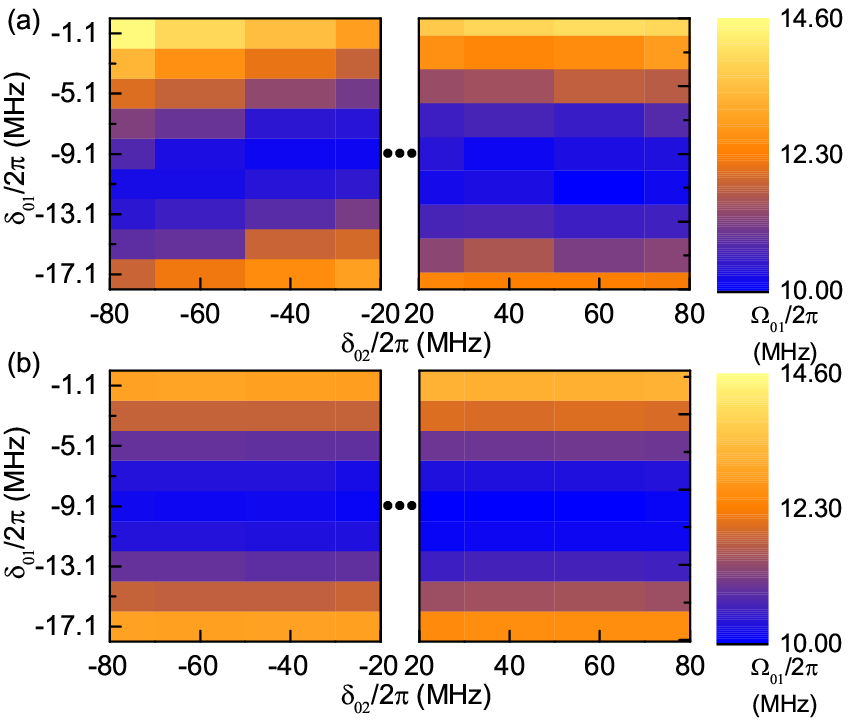}
	\caption{\label{fig:figs2} Rabi oscillation frequency of the 0-1 transition $\Omega_{01}$ versus the detuning of the 0-2 two-photon drive  $\delta_{02}$ and the detuning of the 0-1 drive $\delta_{01}$ shown for (a) experimental observations and (b) simulations.}
\end{figure}

The combined ac-Stark and Bloch-Siegert shifts in the transition frequencies are determined experimentally by the following procedure. First, a 0-2 two-photon drive frequency, which is detuned from the resonant transition frequency, is turned on. This  relatively strong drive field, which has the same amplitude as in the Walsh-Hadamard gate experiments, induces shifts on the energy levels of the device. Next, a weak drive tone is turned on to induce Rabi oscillations in the 0-1 transition, such that the oscillation frequency is comparable to the shifts. Afterwards, both the 0-2 two-photon  and 0-1 drive frequencies are swept and resonant Rabi oscillations of the 0-1 transition is sought after. In the two-photon drive sweeps, we avoid the near-resonance region of the 0-2 two-photon transition because of the non-negligible population oscillation between states 0 and 2. Figure~\ref{fig:figs2}(a) shows the measured Rabi oscillation frequency $\Omega_{01}$ of the 0-1 transition versus the 0-2 two-photon detuning $\delta_{02}$ and the 0-1 detuning $\delta_{01}$. Experimentally, the observed resonant Rabi oscillations indicate a total shift of $2\pi\times9.4$ MHz in the 0-1 transition frequency. The simulation results shown in Fig.~\ref{fig:figs2}(b) are in good agreement with the experimental data.

\section{Evolution with an effective time-dependent Hamiltonian that includes driving-induced corrections}

The relatively strong drive applied to induce the 0-2 photon transition leads to energy level shifts comparable to the driving strengths of the induced dynamics. Therefore it is important to take into account these level shifts accurately in the experiment. During a control pulse, the transition frequencies change, following changes in the driving amplitude. Intuitively, it is expected that the driving frequency has to be dynamically adjusted to follow this change in frequency. In this section we discuss this effect formally in terms of an effective Hamiltonian. In this effective Hamiltonian, the energy eigenstates are the dressed states of the system and the driving terms are represented in terms of their action in the dressed basis.

The effective Hamiltonian is written as
\begin{equation}
\bar{H}(t)=H_{0}(p(t))+H_{\t{d}}(t). \label{eq:ht}  
\end{equation}
The first terms is written explicitly in terms of a parameter $p(t)$. This parameter stands for the amplitude of the strong driving control which induces level shifts. This Hamiltonian is given by
\begin{equation}
H_0(p(t))= \sum_{j=0} \hbar\omega_j(p(t))\ket{j(p(t))}\bra{j(p(t))}, \\
\end{equation}
where $\hbar\omega_j(p(t))$ and $\ket{j(p(t))}$ are, respectively, the shifted levels and dressed states, both of which depend on $p$; at $p=0$ they are the bare levels and states, respectively. The second term of the effective Hamiltonian Eq.~(\ref{eq:ht}) is 
\begin{eqnarray}
H_\t{d}(t)&=& \sum_{\substack{j<k}} A_{jk}(t)  \big[ \ket{j(p(t))}\bra{k(p(t))} + \nonumber
\\&&\ket{k(p(t))}\bra{j(p(t))} \big].
\end{eqnarray}
This terms describes transitions between the dressed eigenstates $\ket{j(p(t))}$. The coefficients  $A_{jk}(t)$ represent the controls associated with the different transitions. We emphasize that these are effective driving terms for these transitions. In particular, for the two photon transitions these terms depend on the square of the applied voltage to the qutrit. 

We consider evolution in a rotating frame described by the unitary operator
\begin{equation}
U(t) = \sum_j\exp(i\theta_j) \ket{j(p(t))}\bra{j(p(t))},
\label{eq:Uframe}
\end{equation}
where $\theta_j (t) = \int_{0}^t\omega_j(p(t^\prime)) \t{d}t^\prime$. In this rotating frame, the matrix elements of the new frame Hamiltonian $\tilde{H}$ between states $m$ and $n$ are
\begin{eqnarray}
\tilde{H}_{mn}(t) &=& \bra{ m(p(t))}i\hbar \dot{U}(t)U^\dagger(t) \ket{n(p(t))} +  \nonumber \\
&&\bra{m(p(t))} U(t) \bar{H}(t) U(t)^\dagger  \ket{n(p(t))}.
\label{eq:framerot}
\end{eqnarray}
The driving amplitudes are 
\begin{equation}
A_{mn}(t) = \bar{A}_{mn} (t) \cos(\theta_{mn}(t) + \varphi_{mn}(t)),
\end{equation}
where $\bar{A}_{mn} (t)$ and $\varphi_{mn}(t))$ are a slowly varying amplitude and phase respectively, and $\theta_{mn}(t) = \theta_m(t) - \theta_n(t)$. With the rotating wave approximation, the Hamiltonian matrix elements in this frame become

\begin{eqnarray}
\tilde{H}_{mn}(t) &=& \frac{\bar{A}_{mn}(t)}{2}  \exp(i\varphi_{nm}(t))
\\
&&+ i\hbar \dot{p} \bigg[   \bra{m(p(t))} \left( \frac{\partial}{\partial p} \ket{n(p(t))} \right) \nonumber
\\
&&+ \left( \frac{\partial}{\partial p} \bra{m(p(t))} \right)\ket{n(p(t))} \exp(i\theta_{nm}(t)) \bigg]. \nonumber
\label{eq:heff}
\end{eqnarray}
Note that the first term is the usual off-diagonal driving element, whereas the next terms are related to the time dependence of the dressed states.

The Hamiltonian~\ref{eq:heff} is expressed in a \emph{time-dependent} basis. The state of the system at any time can be written as the superposition $\sum_n c_n(t) \ket{n(t)}$. When the Schr\"{o}dinger equation is written for this wavefunction, the coefficients can be easily shown to evolve in time as 
\begin{equation}
i\hbar \dot{c}_m(t) = \sum_n \left( \tilde{H}_{mn}(t) - h_{mn}(t)\right) c_n(t),
\end{equation}
where $h_{mn}(t)$ is a correction term given by $h_{mn}(t) = i\hbar \dot{p}  \bra{m(p(t))} \left( \frac{\partial}{\partial p} \ket{n(p(t))} \right)$. The evolution of these coefficients is described by the new Hamiltonian
\begin{eqnarray}
\tilde{H}'_{mn}(t) &=& \frac{\bar{A}_{mn}(t)}{2}  \exp(i\varphi_{nm}(t))
\label{eq:heffcorrected} \\
&&+ i\hbar \dot{p} \left( \frac{\partial}{\partial p} \bra{m(p(t))} \right)\ket{n(p(t))} \exp(i\theta_{nm}(t)). \nonumber
\end{eqnarray}
It is important to note that the Hamiltonian~\ref{eq:heffcorrected} should be understood as describing the evolution of the coefficients in a time-dependent basis. However, the states at the beginning and end of a pulse, where $p=0$, coincide and are the same as the bare states.

Next we comment on the structure of the Hamiltonian~\ref{eq:heffcorrected}. While the first term is the usual driving term, the second term represents corrections due to the time dependence of the energy eigenstates. These corrections can be neglected on account of the following arguments. The off-diagonal terms ($m \neq n$) are neglected due to the fast oscillations in $\exp(i\theta_{nm}(t))$, assuming that the rise and fall of the control pulses, reflected in $\dot{p}(t)$, are slow. The diagonal part can be neglected on account of the fact that the derivatives of the states are small when the perturbation of the states due to driving are small and, in addition, due to the fact that the rising and falling parts of the pulse lead to corrections of different signs.

The discussion in this section gives a full justification for the use of driving pulses that are applied at a time-dependent frequency, matching the instantaneous transition frequencies between dressed states. With this condition, the evolution can be described by a fully off-diagonal Hamiltonian. The transformation between the rotating frame and the lab frame is given by~\ref{eq:Uframe} and contains phase corrections due to the level shifts induced by the driving.

\section{Numerical simulations of the system dynamics}
In this section we describe the numerical simulations used to produce the simulation results presented in the main text. For a voltage drive signal $V_\t{g}(t)$, the simulations use the Hamiltonian $H|_{V_\t{g} = 0} + \frac{\partial H}{\partial V_\t{g} }|_{V_\t{g} = 0} V_\t{g}(t)$. Here, $H$ is the circuit Hamiltonian [Eq.~(\ref{eq:ham})] in the absence of driving and truncated to the lowest few energy levels, with the number of levels used as a variable parameter in the simulations to understand the role of higher levels. The time-dependent function $V_\t{g}(t)$ is the driving voltage applied to the qutrit control gate.

The parameters in the circuit Hamiltonian (capacitance matrix, Josephson junction energies) are first adjusted based on a fit of the system spectroscopy. This fit gives excellent agreement for the first two transition frequencies versus flux around the qutrit symmetry point at half a flux quantum in the loop. While the parameters for the circuit are adjusted based on three-level spectroscopy, we expect that this model may not be a sufficiently close representation of the measured system due to limitations of the circuit model and the presence of the resonator. Therefore, we `calibrate' the drives, in a procedure that closely follows the experiment. This is done as follows. Voltages are applied to frequencies 01 and 12, and the Rabi frequency per applied voltage is characterized by observing the oscillations in the population. Similarly, with a tone applied at half the frequency of the 0-2 transition, we characterize both the Rabi strength of the 0-2 Rabi oscillations and the detuning. We find that the ratio of the Rabi strength and detuning for 0-2 closely matches the experimental results. Additionally, the ac voltages needed to obtain Rabi frequencies equal to the experimental values for the 0-1, 1-2 and 0-2 transitions are in good agreement with the voltages applied in the experiment and the microwave transmission in the setup.

With this `simulation calibration' performed, we next simulate the quantum time dynamics for two cases. In the first case, no decoherence is included in the evolution of the system, and hence the Schr\"odinger equation is used to obtain the dynamics. In the second case, the simulations assume that the evolution with decoherence is described by the full Linblad form
\begin{eqnarray}
\dot{\rho} &=& -i[H,\rho]+\sum_{i<j} \Gamma_{ij}\mathcal{L}[\sigma_{ij}^+]\rho + \sum_{i>j} \Gamma_{ij}\mathcal{L}[\sigma_{ij}^-]\rho \nonumber \\
&&
+\sum_{i<j}\Gamma_{ij}^\t{R}\mathcal{L}[\sigma_{ij}^{z}]\rho ,
\label{eq:simLinblad}
\end{eqnarray}
where $\sigma_{ij}^{z} = \ket{i} \bra{i} - \ket{j} \bra{i}$ and all the other symbols are as defined in subsection~\ref{subsec:mlc}. We note that here we used a master equation to model not only relaxation/excitation but also pure dephasing, which is not fully consistent with the coherence functions being Gaussian but is the standard approach used in the literature and is expected to give a good approximation to the decoherence dynamics. Treating pure dephasing separately and including 1/f noise for a driven system would result in unnecessarily complicated simulation models. One additional important point is that we only include the decoherence processes corresponding to the qutrit states (0, 1, 2), while the simulation is carried out including levels outside this subspace; we expect this to be a reasonable approximation, given that the population of higher levels remains negligible throughout the dynamics. Including these higher levels is important for properly taking into account the level shifts. For both cases, the evolution is compared against the ideal evolution, with measures of fidelity identical to those in the experiment.

\section{Quantum state Tomography}
The tomography analyzer pulses are two-state unitaries and  consist of rotations around the x and y axes of the Bloch sphere for the 0-1 and 1-2 transitions. These rotations can be denoted by $R_{\mathbf{n}}^{01,12}(\theta)$ where $\mathbf{n}$ is the rotation axis and $\theta$ is the rotation angle. In the experiments, the diagonal part $U_\t{d}$ of the decomposed Walsh-Hadamard gate is absorbed into the tomography pulses by shifting the drive phases of these pulses. In the rotating frame, the modified tomography  pulses,  given by $\bar{u}_k=U_d^\dagger u_k U_d$ can be considered as rotations around a new axis $\mathbf{n}'$. For the 0-1 transition, the  rotation axis  $\mathbf{n}' = (n'_x, n'_y, 0)$ is related to $\mathbf{n} = (n_x, n_y, 0)$ by
\begin{equation}
\begin{pmatrix}
n'_x \\
n'_y \\
0
\end{pmatrix} = 
\begin{pmatrix}
\cos(\phi_0-\phi_1) & \sin(\phi_0-\phi_1) & 0\\
-\sin(\phi_0-\phi_1) & \cos(\phi_0-\phi_1) & 0\\
0 & 0 & 1
\end{pmatrix}
\begin{pmatrix}
n_x \\
n_y \\
0
\end{pmatrix}.
\end{equation}
The angles $\phi_0$ and $\phi_1$ are the phases in $U_d$. A similar relation can be derived for other transitions.

The density matrix of a given state is reconstructed with these new analyzer pulses. For this purpose we employ maximum likelihood estimation (MLE) following James \emph{et. al.}~\cite{SIjames_measurement_2001}. A likelihood function is introduced that measures the closeness of the generated physical density matrix to the experimentally measured one. For a qutrit we adopt a similar approach. The expression for a physical density matrix is given by
\begin{equation}
\rho(t)=\frac{T^\dagger(t)T(t)}{\t{Tr}[T^\dagger(t)T(t)]},
\end{equation}
with optimization parameters
\begin{equation}
T(t)=
\begin{pmatrix}
t_1&0&0\\
t_4+ it_5&t_2&0\\
t_8 +it_9&t_6+ it_8& t3
\end{pmatrix}.
\end{equation}
The density matrix is reconstructed by minimizing the likelihood function $L=\sum_{k}|V_k-\t{Tr}[ \rho\tilde{u}_k V_\t{h} \tilde{u}_k^\dagger]|$, where $V_h$ is the readout homodyne voltage operator and $V_k$ are the measured voltage values. 

\section{Quantum process tomography}
The quantum process tomography procedure is implemented as follows. Based on a nominal state $\rho_\t{n}$ (which is either the qutrit thermal state or the state obtained after the cooling process), we generate, by applying unitaries $p_j$ ($j=\overline{0,8}$), nine different input states $\rho_{\t{in}, j} = p_j \rho_\t{n} p_j^\dag$. The measurement procedure of the qubit is based on standard circuit QED and is modelled by the measurement of the operator $V_\t{h} = V_{\t{h}0} \ket{0}\bra{0} + V_{\t{h}1} \ket{1}\bra{1} + V_{\t{h}2} \ket{2}\bra{2}$. By inserting each of the nine tomography pulses $u_i$ ($i=\overline{0,8}$) prior to the measurement, one generates nine different measurement operators $M_i = u_i^\dag V_\t{h} u_i$. 

The quantum process is described by the transformation of an input state $\rho_\t{in}$ to an output state $\rho_\t{out}$, given by $\rho_\t{out} (\rho_\t{in}) = \sum_{m, n = \overline{0,8}} \chi_{m n} \lambda_m \rho_\t{in} \lambda_n^\dag $. Here $\chi_{m n}$ are the elements of the process matrix $\chi$ and $\lambda_m$ ($m = \overline{0,8}$) are matrices with the property $\t{Tr} (\lambda_m \lambda_n^\dag )= d\,\delta_{mn} $, with $d = 3$ the dimension of the qutrit subspace~\cite{SIjames_measurement_2001}. With this parameterization, the average value for the measurement with tomography setting $i$ and with preparation setting $j$ is given by $m_{ij} =\sum_{m, n} \chi_{m n} \t{Tr} (M_i  \lambda_m p_j \lambda_n^\dag )$.

The process matrix $\chi$ is a positive matrix of trace 1, parameterized using the Cholesky decomposition. We use a MLE method to find the parameters in the $\chi$ matrix based on minimizing the quantity $\sum_{i,j = \overline{0,8}} (m_{ij} - m_{ij}^\t{exp})^2$, where $ m_{ij}^\t{exp}$ are the experimental results for the measurement $i$ with preparation $j$. The ideal process matrix corresponding to an unitary $U$ is given by $\chi_{mn}^\t{ideal} = e_m\, e_n^*$, where we decomposed the unitary as $U = \sum_m e_m \lambda_m$ in terms of the complete set ${\lambda_m}$. The process fidelity is given by $\t{Tr} \left( \sqrt{ \sqrt{\chi^\t{ideal}} \chi \sqrt{\chi^\t{ideal}}} \right)$~\cite{SInielsen2010quantum}.  
\\


%

\end{document}